\documentclass{emulateapj}

\slugcomment{Accepted for publication in ApJL, 2011 May 6}

\shorttitle{FIR counterparts and dust content of LAEs at z$\lesssim$1.0}
\shortauthors{Oteo et al.}

\usepackage{natbib}

\begin{document}


\title{FIR measurements of Ly-$\alpha$ emitters at z$\lesssim$1.0: dust 
attenuation from PACS-\emph{Herschel}}

\author{I. Oteo\altaffilmark{1,2}, A. Bongiovanni\altaffilmark{1,2}, A. M. P\'erez Garc\'{\i}a\altaffilmark{1,2}, J. Cepa\altaffilmark{2,1}, 
A. Ederoclite\altaffilmark{1,2}, M. S\'anchez-Portal\altaffilmark{3}, I. Pintos-Castro\altaffilmark{1,2}, D. Lutz\altaffilmark{4} S. Berta\altaffilmark{4}, E. Le Floc'h\altaffilmark{5}, B. Magnelli\altaffilmark{4}, P. Popesso\altaffilmark{4}, F. Pozzi\altaffilmark{6}, L. Riguccini\altaffilmark{7}, B. Altieri\altaffilmark{3}, P. Andreani\altaffilmark{8}, H. Aussel\altaffilmark{9}, A. Cimatti\altaffilmark{10}, E. Daddi\altaffilmark{6}, D. Elbaz\altaffilmark{9}, N. F\"orster Schreiber\altaffilmark{4}, R. Genzel\altaffilmark{4}, R. Maiolino\altaffilmark{11}, A. Poglitsch\altaffilmark{4}, E. Sturm\altaffilmark{4}, L. Tacconi\altaffilmark{4}, I. Valtchanov\altaffilmark{3}}

\altaffiltext{1}{Instituto de Astrof{\'i}sica de Canarias (IAC), E-38200 La Laguna, Tenerife, Spain}
\altaffiltext{2}{Departamento de Astrof{\'i}sica, Universidad de La Laguna (ULL), E-38205 La Laguna, Tenerife, Spain}
\altaffiltext{3}{Herschel Science Centre (ESAC). Villafranca del Castillo, Spain}
\altaffiltext{4}{Max-Planck-Institut f\"{u}r Extraterrestrische Physik (MPE), Postfach 1312, 85741 Garching, Germany}
\altaffiltext{5}{Institute for Astronomy, University of Hawaii, 2680 Woodlawn Drive, Honolulu, HI 96822, USA}
\altaffiltext{6}{INAF - Osservatorio Astronomico di Bologna, via Ranzani 1, I-40127 Bologna, Italy}
\altaffiltext{7}{Laboratoire AIM, CEA/DSM-CNRS-Universit{\'e} Paris Diderot, IRFU/Service d'Astrophysique, B\^at.709, CEA-Saclay, 91191 Gif-sur-Yvette Cedex, France}
\altaffiltext{8}{ESO, Karl-Schwarzchild-Str. 2, D-85748 Garching, Germany}
\altaffiltext{9}{Commissariat \`a l'\'Energie Atomique (CEA-SAp) Saclay, France}
\altaffiltext{10}{Dipartimento di Astronomia, Universit\`a di Bologna, Via Ranzani 1, 40127 Bologna, Italy}
\altaffiltext{11}{INAF - Osservatorio Astronomico di Roma, via di Frascati 33, 00040 Monte Porzio Catone, Italy}

\begin{abstract}
One remaining open question regarding the physical properties of Ly$\alpha$ emitters (LAEs) is their dust content and its evolution with redshift. The variety of results is large and with those reported by now is difficult to establish clear relations between dust, other fundamental parameters of galaxies (star-formation rate, metallicity or age) and redshift. In this Letter, we report \emph{Herschel} PACS-100$\mu$m, PACS-160$\mu$m and \emph{Spitzer} MIPS-24$\mu$m detections of a sample of spectroscopically GALEX-selected LAEs at z$\sim$0.3 and $\sim$1.0. Five out of ten and one out of two LAEs are detected in, at least, one PACS band at z$\sim$0.3 and $\sim$1.0, respectively. These measurements have a great importance given that they allow us to quantify, for the first time, the dust content of LAEs from direct FIR observations. MIPS-24$\mu$m detections allow us to determine IR properties of the PACS-undetected LAEs. We obtain that mid-IR/FIR detected star-forming (SF) LAEs at z$\sim$0.3 have dust content within 0.75$\lesssim$ $A_{1200\AA}$ $\lesssim$2.0, with a median value of A$_{1200\textrm{\AA}}$$\sim$1.1. This range broadens out to 0.75$\lesssim$ $A_{1200\AA}$ $\lesssim$2.5 when considering those LAEs at z$\sim$1.0. Only one SF LAE is undetected both in MIPS-24$\mu$m and PACS, with $A_{1200\AA}$ $\lesssim$0.75. These results seem to be larger than those reported for high-redshift LAEs and, therefore, although an evolutionary trend is not clearly seen, it could point out that low-redshift LAEs are dustier than high-redshift ones. However, the diverse methods used could introduce a systematic offset in the results.
\end{abstract}



\section{Introduction}

Ly$\alpha$ emitters (LAEs) have been proved to be of great importance in the study of the formation and evolution of galaxies through cosmic time. At 2$\lesssim$z$\lesssim$7, a large number of candidates have been found, mainly via the narrow or intermediate band technique in the optical/NIR wavelength range \citep{Ouchi2008, Cowiehu1998, Gronwall2007, Guaita2010, Shioya2009, Murayama2007, Nilsson2009, Fujita2003, Bongiovanni2010}. At z$\lesssim$2, the LAEs catalogued up to now have been found by spectroscopic surveys using the \emph{Galaxy Evolution Explorer} \citep[GALEX,][]{Martin2005}. The selection technique is based on looking for a Ly$\alpha$ line in the UV spectra of objects with a measured UV continuum \citep{Deharveng2008,Cowie2010}. 

Despite of the large number of LAE candidates found at different redshifts, their physical properties and their evolution with cosmic time are not completely understood yet. Moreover, the relation between LAEs and other kind of high redshift galaxies, such as Lyman break galaxies (LBGs) or distant red galaxies (DRGs), remains unclear. Some studies have suggested that the stellar populations of LAEs exhibit a double nature, some being old and massive and other young and less massive \citep{Gawiser2006,Lai2008,Ono2010}. However, recent studies have proved that stellar populations of LAEs vary continuously and smoothly over a wide range, conforming a big zoo of galaxies \citep[Oteo et al. 2011, submitted;][]{Nilsson2011}.

One remaining open question is whether LAEs are primitive and dust-free objects, evolved and dusty, or present a wide range of dust attenuation. The Ly$\alpha$ line has a resonant nature and, therefore, the probability of being absorbed by the interstellar medium (ISM) even before leaving out the host galaxy is quite high. Due to that, it was initially thought that LAEs found via a narrow band technique should be young and dust free objects, allowing the Ly$\alpha$ line to appear in emission. However, \cite{Neufeld1991} proposed a scenario where the Ly$\alpha$ equivalent width could be enhanced by the presence of dust when it resides in cold and neutral clumps located within an ionized ISM. Actually, some studies \citep{Finkelstein2009, Finkelstein2009d, Blanc2010} have observationally confirmed this hypothesis, opening the possibility of finding the Ly$\alpha$ emission in non primordial galaxies at different redshifts. The dust content in LAEs has been studied in a wide range of redshifts and different results have been found, from dust-free objects to low, moderate, and even high dust content. In Table \ref{dust_sum}, some results on the study of dust attenuation in LAEs at different redshifts are summarized. 

Most previous works have studied the dust attenuation in LAEs by fitting the UV and optical rest-frame spectral energy distribution (SED) with stellar population templates, mainly \cite{Bruzual2003} (BC03) ones, and assuming the dust attenuation law of \cite{Calzetti1997}. However, both dust and old stars can redden the rest-frame integrated UV continuum, varying the expected colors from the dust-free scenario. UV and optical photons which are absorbed by dust are reemitted in the FIR region. Therefore, the most accurate way to study the dust properties of high redshift galaxies is observing the dust emission in the FIR. \cite{Finkelstein2009a} and \cite{Dayal2010} studied the expected detection of dust emission from LAEs at z$\gtrsim$4 and z$\sim$6, respectively, finding that a large percentage of them at those redshifts should be detected in only a few hours with ALMA. \cite{Finkelstein2009a} also found that LBGs are 60\% more likely to be detected than LAEs, meaning that they are dustier objects and showing a possible difference between the two populations. However, some recent studies did not find FIR counterparts for LBGs z$\sim$3, characterizing their FIR SED only with stacked photometry \citep{Magdis2010_PACS, Rigopoulou2010}. For LAEs, \cite{Bongiovanni2010} reported the detection in the FIR of 3 LAEs at z$\sim$2.3 selected from the ALHAMBRA survey \citep{Moles2008}, being their optical to mid-IR SED well fitted by AGN-like templates of \cite{Polletta2007}. Oteo et al. (2011a, submitted), also found FIR counterparts for 5 spectroscopically confirmed LAEs at z$\sim$2.5. In this case, according to their optical spectra, only one is an AGN, and the other four are star-forming (SF) LAEs. All of these FIR-detected LAEs resulted to be (U)LIRGs candidates and, therefore, red and dusty objects, despite having the Ly$\alpha$ line in emission and with large equivalent widths. \cite{Chapman2005} also found evidences of the Ly$\alpha$ emission in a sample of sub-mm galaxies detected in 850$\mu$m, being most of them (U)LIRGs candidates too.

Until now, few FIR counterparts for LAEs at z$\lesssim$2 have been reported, and for those LAEs with FIR counterparts, an extensive analysis of their FIR properties has not been done. In this letter, we report the FIR detections of a sample of LAEs at z$\sim$0.3 and z$\sim$1.0 by using data coming from Photodetector Array Camera and Spectrometer \citep[PACS,][]{Poglitsch2010} on board \emph{Herschel Space Observatory}. These FIR measurements, combined with UV information, enable us to determine their dust content in an accurate way without the uncertainties coming from the properties of the dust itself, as its emission in the FIR, missing in the optical methods.

This Letter is organized as follows: In Sections \ref{galex} and \ref{pacs} we present the LAE sample and the FIR data used, respectively. In Section \ref{counterparts} we report the FIR counterparts of our LAEs and in Section \ref{dust_content} we study their dust attenuation, comparing them with previous results. In Section \ref{conclusion}, the main conclusions of our work are described.

Throughout this work we assume a flat universe with $(\Omega_m, \Omega_\Lambda, h_0)=(0.3, 0.7, 0.7)$, and all magnitudes are listed in the AB system \citep{Oke1983}.

\section{LAE sample}\label{galex}

\begin{figure*}[!!!t]
   \centering
   \includegraphics[width=0.49\textwidth]{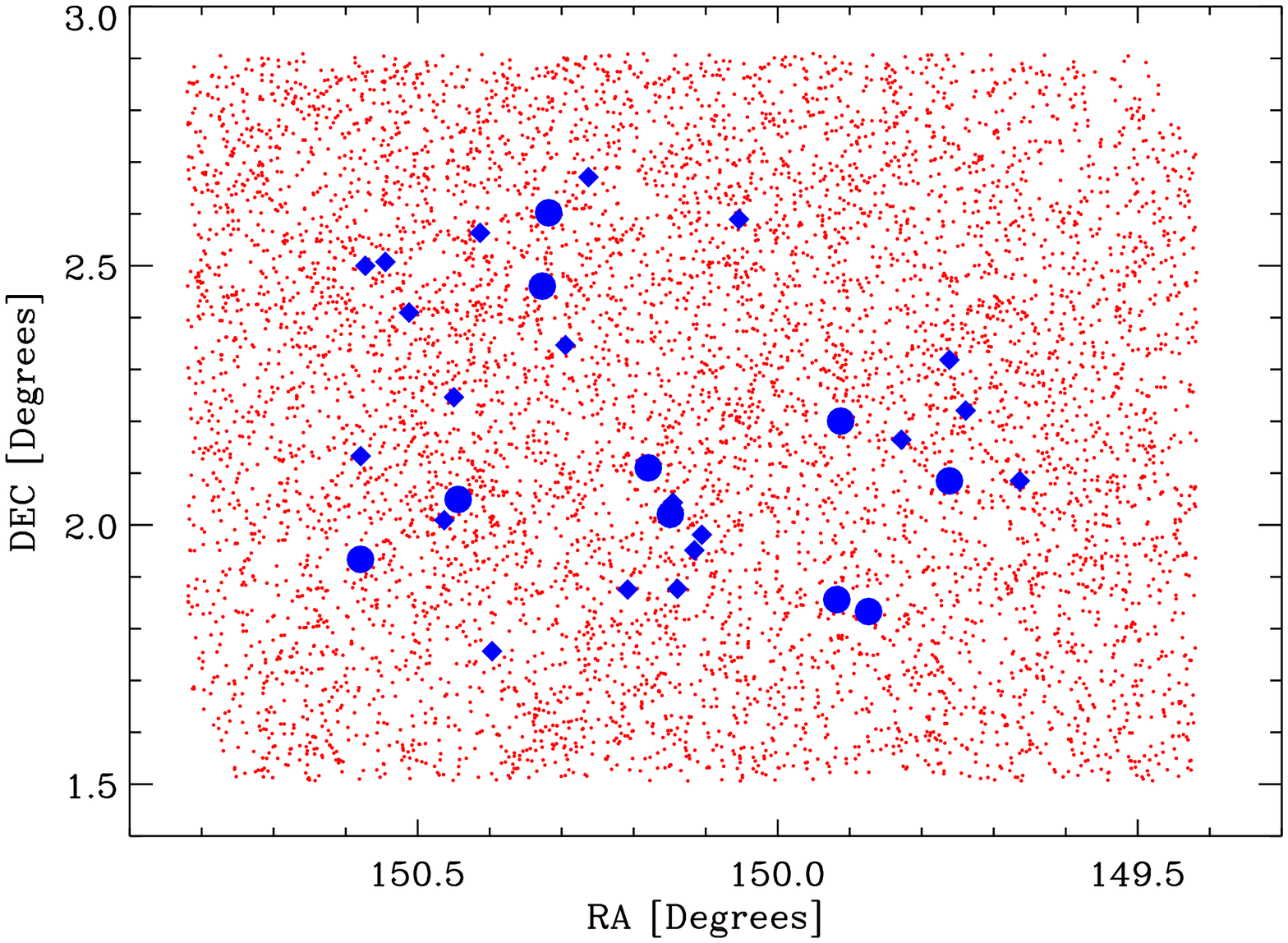}
      \includegraphics[width=0.49\textwidth]{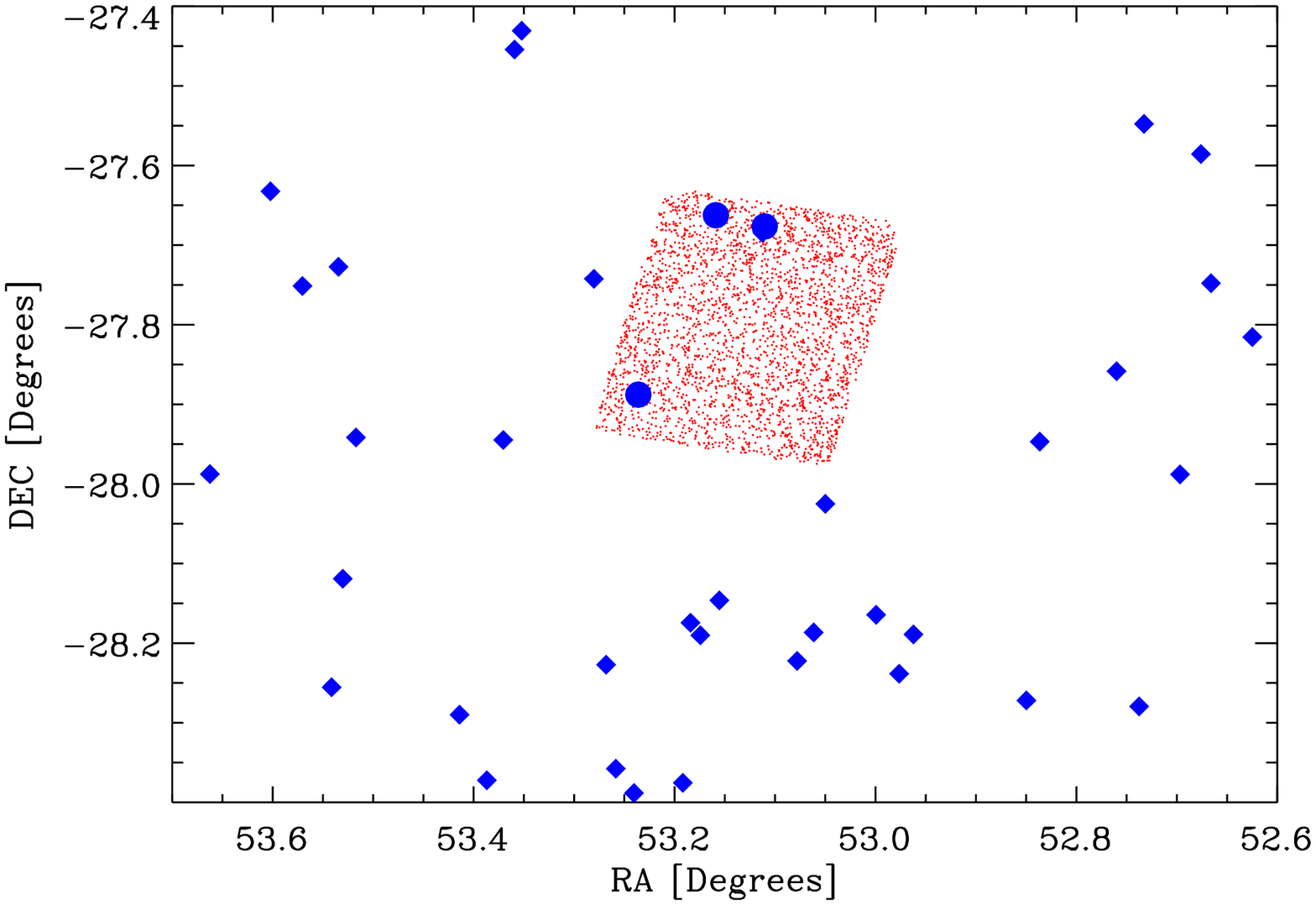}
\caption{Spatial distribution maps of the GALEX-LAEs (both AGNs and SF) and PACS sources. Blue diamonds and blue dots are PACS-undetected and PACS-detected GALEX-LAEs, respectively. Red dots are PACS detections with 24$\mu$m priors. \emph{Left:} COSMOS field. \emph{Right:} GOODS-South fields. It can be seen that all the GALEX-LAEs in COSMOS are within the covered area of PACS. In GOODS-South, only four GALEX-LAEs are within the PACS imprint. Note in the left plot that one of the PACS-undetected sources is so close to a PACS-detected one that the point is not clearly seen.}             
         \label{mapillas}
   \end{figure*}

In our analysis, we use an LAE sample selected from \cite{Cowie2010}, where the spectra, taken with GALEX NUV and FUV grisms, were selected from the Multimission Archive at STScI (MAST). \cite{Cowie2010} used an automatic search procedure to look for emission lines in the spectra of objects with a measured UV continuum. For each source, a polynomial curve was used for constraining the continuum, and a gaussian profile fit to look for emission lines. Given that the spectra become very noisy at the edges of the spectral range in each channel, only those objects which have the Ly$\alpha$ emission in the wavelength ranges 1452.5-1750 \AA$\,$ and 2006-2735 \AA$\,$ for the FUV and NUV grisms, respectively, were selected. In the redshift space, this means that the sources are located within z=0.195-0.44 (z$\sim$0.3 sample) and z=0.65-1.25 (z$\sim$1.0 sample) in the FUV and NUV spectra, respectively. Both samples are nearly complete to an NUV limiting magnitude of 21.5 and have Ly$\alpha$ luminosity limits of 5$\cdot 10^{41}$ and 1$\cdot10^{43}$ erg cm$^{-1}$s$^{-1}$ at z$\sim$0.3 and $\sim$1.0, respectively. LAEs were classified as AGN or star-forming (SF) galaxies according to the width and shape of the Ly$\alpha$ line and the presence or absence of AGN ionization lines in the UV spectra. Figure \ref{mapillas} shows the spatial distribution of GALEX-selected LAEs in both COSMOS and GOODS-South fields.

\section{FIR data}\label{pacs}

GOODS-South and COSMOS fields have been observed with PACS on board \emph{Herschel} in the framework of PACS Evolutionary Probe project (PEP, PI D. Lutz). PEP is the \emph{Herschel} Guaranteed Time Key-Project to obtain the best profit for studying FIR galaxies evolution from \emph{Herschel} instrumentation (Lutz et al. 2011, in prep). For both fields, PACS-100$\mu$m and PACS-160$\mu$m observations are available, with 3$\sigma$ limiting fluxes of (1.1, 5.0) in PACS-100$\mu$m and (2.0, 11.0) in PACS-160$\mu$m in (GOODS-South, COSMOS). PACS-100$\mu$m and PACS-160$\mu$m bands cover the FIR wavelength ranges 80-130$\mu$m and 130-210$\mu$m, respectively. Two kind of catalogues have been built: blind and with MIPS-24$\mu$m position priors by using \cite{Lefloch2009} and \cite{Magnelli2011} catalogs in COSMOS and GOODS-South, respectively. In order to have as much FIR information as possible and more reliability in the identification of the sources, we use the catalogs extracted with MIPS-24$\mu$m priors. Data reduction, catalogs construction and simulations, aimed at deriving completeness, fraction of spurious sources and photometric reliability, are described in Lutz et al. (2011, in prep.).

Figure \ref{mapillas} also shows the PACS-\emph{Herschel} sources extracted with 24$\mu$m positions priors. It can be seen that PACS observations in COSMOS field cover the whole region where GALEX-selected LAEs are located, whereas the surveyed area by PACS in GOODS-South only covers a little portion of GALEX observations. In this sense, 28 and 4 GALEX-selected LAEs are within the PACS surveyed area in COSMOS and GOODS-South, respectively. Table \ref{table_number} summarizes the number of GALEX-selected LAEs within the PACS area in both fields.

\section{LAEs with FIR counterparts}\label{counterparts}

In order to find out how many LAEs have FIR counterparts we performed a match between GALEX-LAEs coordinates and PACS catalogs in both fields. We look for a possible FIR detection within a radius of 2'' around the position of each LAE, since it is the astrometric uncertainty in the position of the sources. With this procedure, we find that a noticeable fraction of LAEs, mostly above 30\%, have PACS-FIR detections in at least one considered PACS band, considering limiting fluxes shown in Sect. \ref{pacs}. In Table \ref{table_number}, the number of PACS detection are also shown, separated by field, redshift and nature of the object (AGN or SF). The percentages of FIR detections are larger in GOODS-South than in COSMOS because of the depth of the observations. If the PACS observation had been in COSMOS as deep as in GOODS-South, a large number of FIR counterparts would have been expected.

\begin{figure*}
   \centering
      \includegraphics[width=0.23\textwidth]{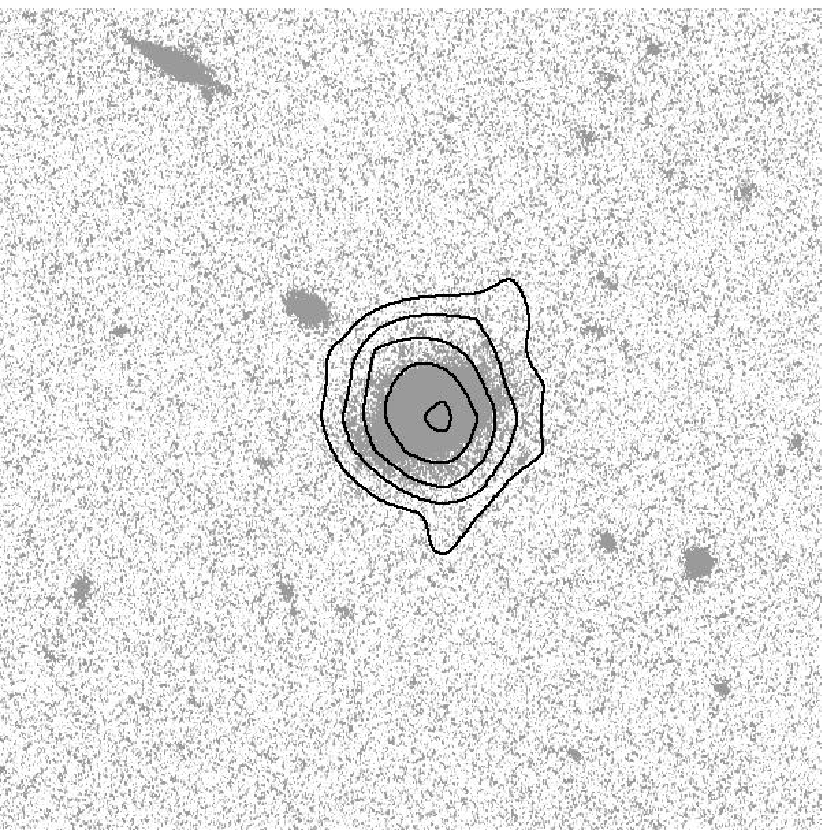}
      \includegraphics[width=0.23\textwidth]{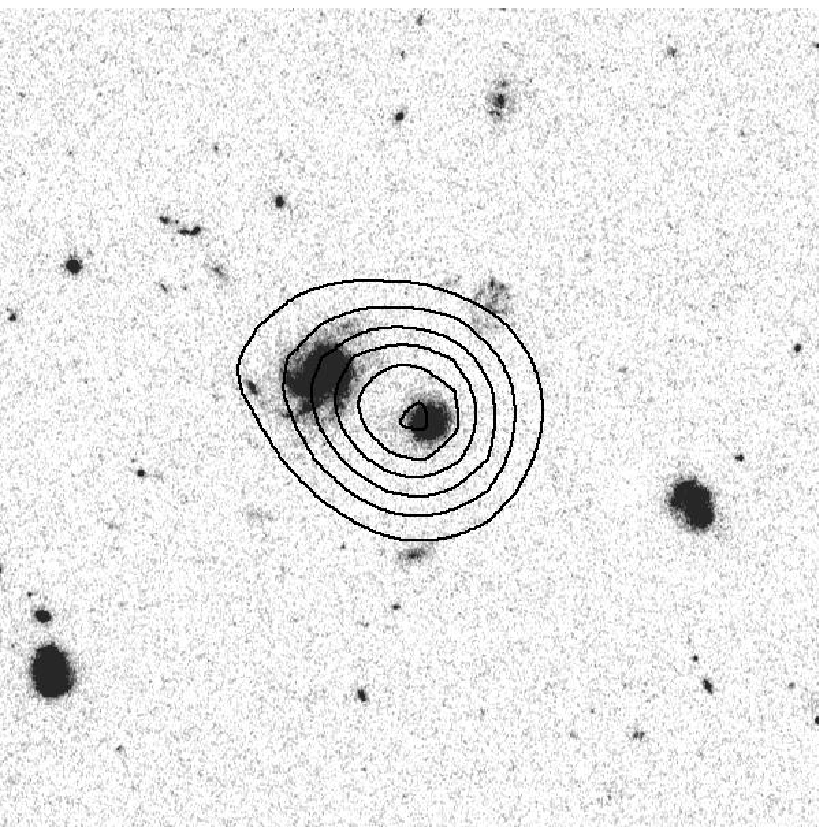}
      \includegraphics[width=0.23\textwidth]{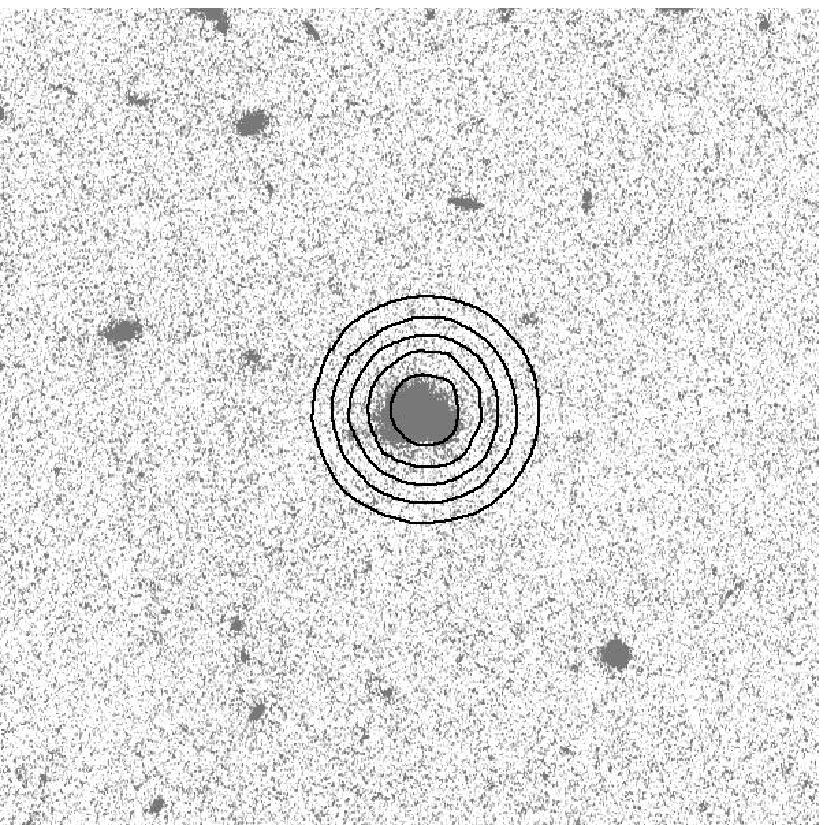}
      \includegraphics[width=0.23\textwidth]{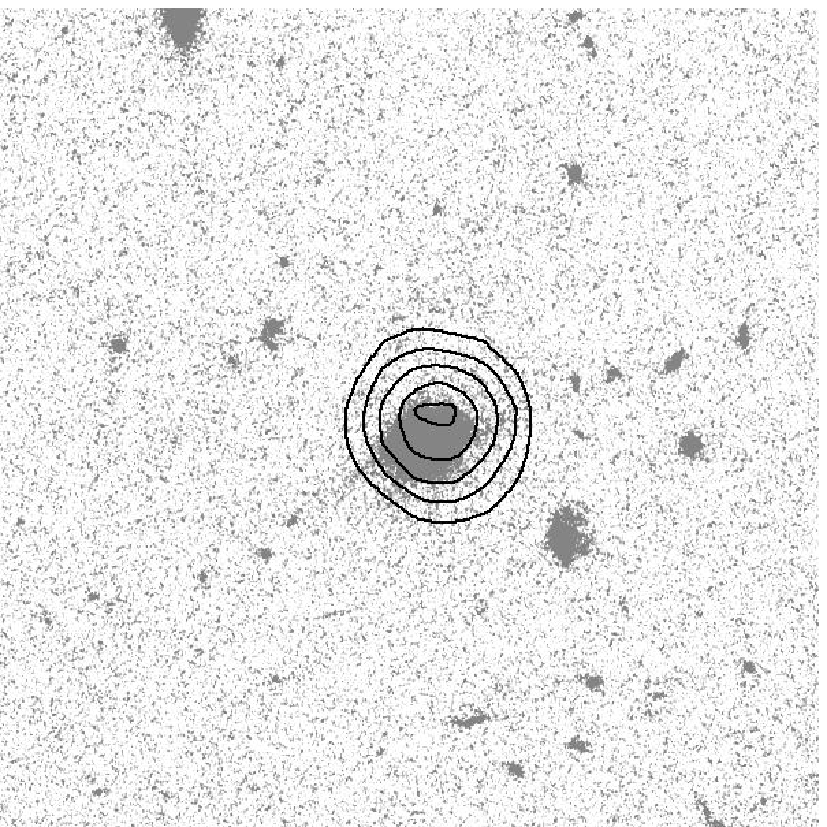}\\
      \includegraphics[width=0.23\textwidth]{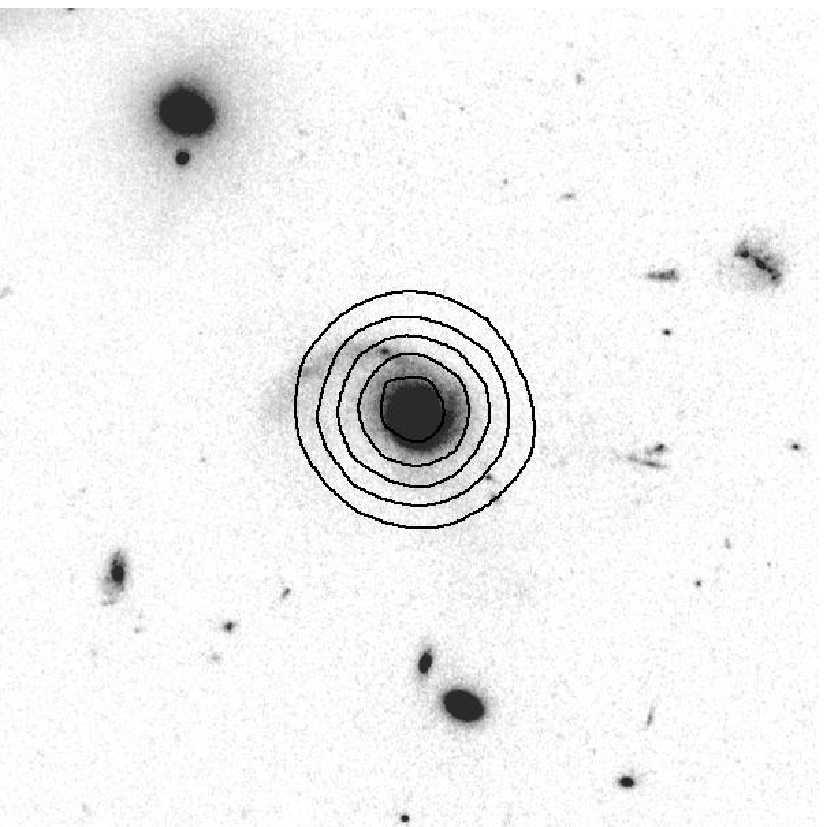}
      \includegraphics[width=0.23\textwidth]{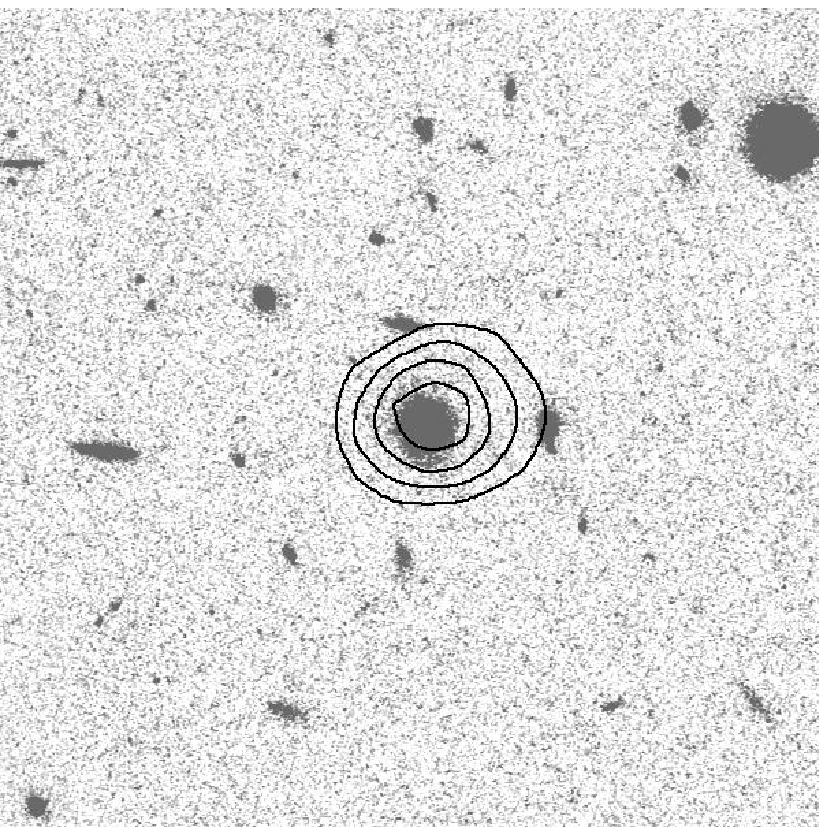}
      \includegraphics[width=0.23\textwidth]{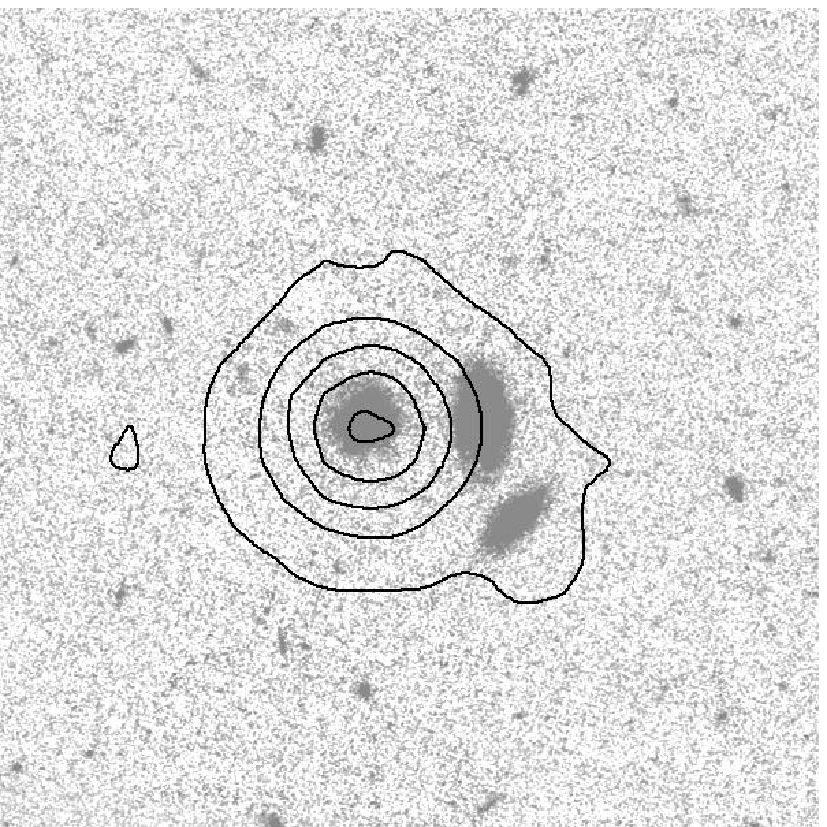}
      \includegraphics[width=0.23\textwidth]{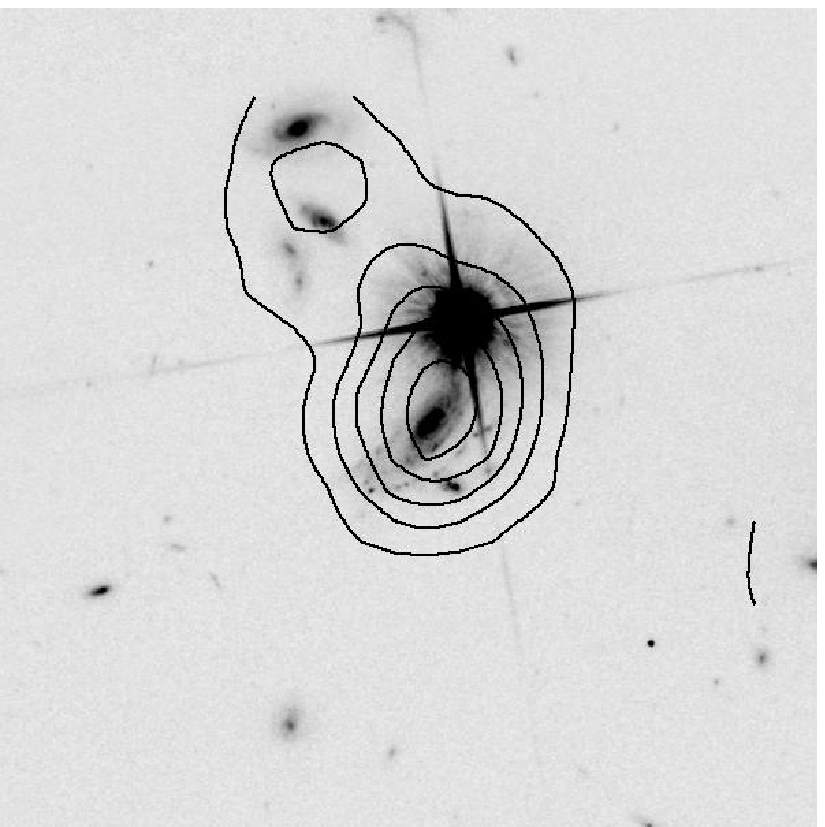}\\
      \includegraphics[width=0.23\textwidth]{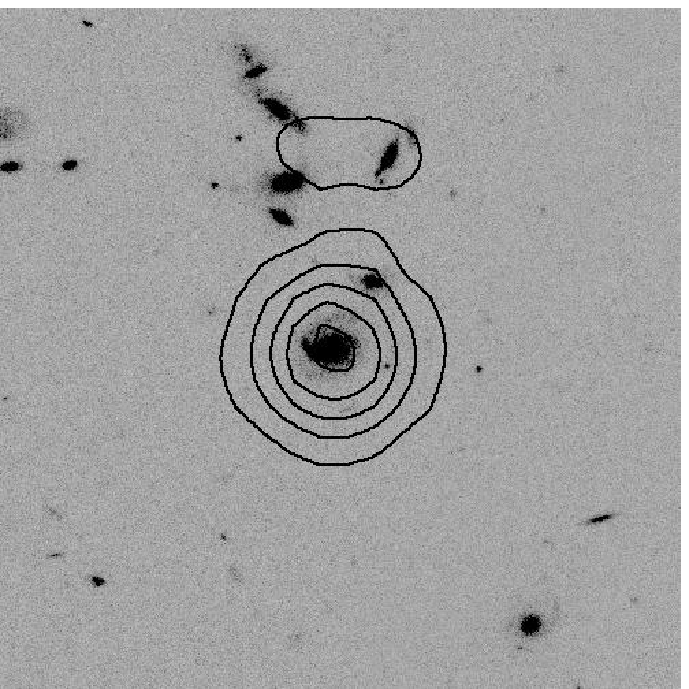}
      \includegraphics[width=0.23\textwidth]{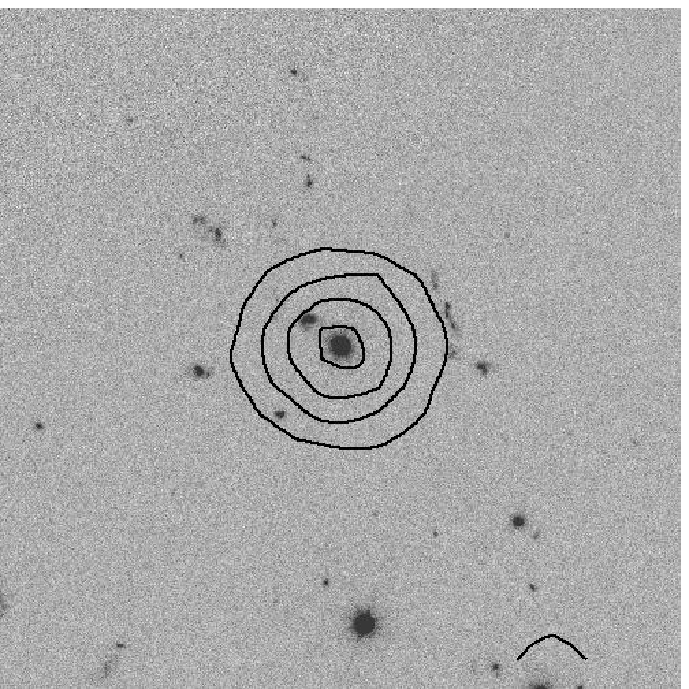}
      \includegraphics[width=0.23\textwidth]{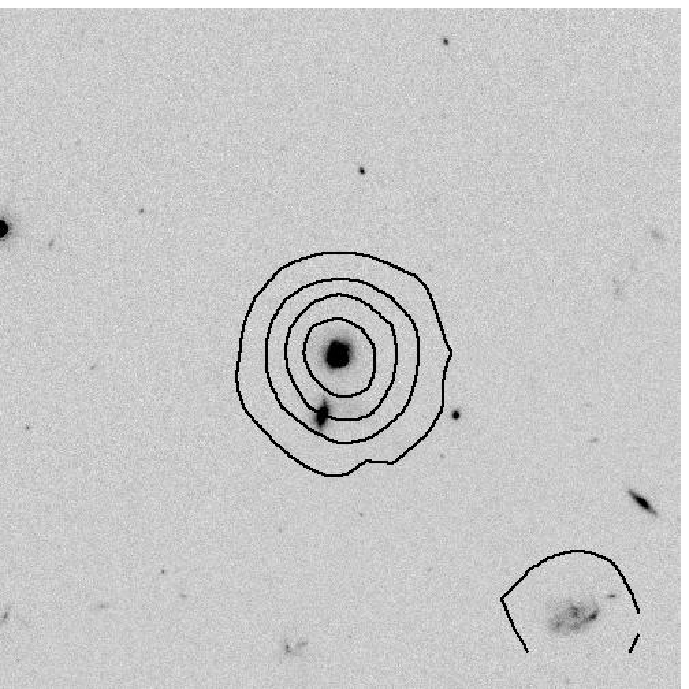}
   \caption{30'' x 30'' ACS-optical cut-outs and MIPS-24$\mu$ flux isocontours, linearly scaled to the maxima. GALEX-LAEs are the objects located in the center of each image. Black curves are flux iso-contours in MIPS-24$\mu$m band, linearly scaled to the peak flux.
              }
           \label{sources}
   \end{figure*}

A source confusion analysis is needed to ensure that each FIR detection corresponds to the matched GALEX-selected LAE. We take 30'' optical cut-outs around the position of each GALEX-selected LAE and overplot the MIPS-24$\mu$m contours in order to verify if the detections in each band is related or not to the same object. We use the MIPS-24$\mu$m contours given that the PACS catalogues are based in 24$\mu$m positions priors, where the PSF of the observations is $\sim$30\% narrower than redder bands and, therefore, the accuracy in the positions is better. It can be seen in Figure \ref{sources} that each FIR detection reported corresponds to the matched GALEX-LAE.

\section{Dust attenuation in star-forming LAEs}\label{dust_content}
   
The ratio between the total IR and UV luminosities, L$_{IR}$ and L$_{UV}$, respectively, is directly related to the dust attenuation in galaxies, and calibrations have been already built \citep{Meurer1999,Buat2005,Buat2007}. Since all the PACS-detected SF GALEX-LAEs also have UV measurements, we can infer their dust attenuation from direct observations of their FIR SED. We use the NUV band given that, on one hand, it is not contaminated by the Ly$\alpha$ emission for those GALEX-LAEs at z$\sim$0.3 and, on the other hand, it is not affected by the Lyman break for those GALEX-LAEs at z$\sim$1.0. L$_{UV}$ can be obtained directly from the observed magnitudes in the NUV band, as in \cite{Buat2005}. L$_{IR}$ can be calculated by using the calibrations given in Oteo et al. (2011a, submitted) between PACS bands and total infrared luminosities:

\begin{equation}\label{cal1}
\log{L_{IR}} = 0.99 \log{L_{100\mu\textrm{m}}} + (0.44 \pm 0.25)
\end{equation}

\begin{equation}\label{cal2}
\log{L_{IR}} = 0.96 \log{L_{160\mu\textrm{m}}} + (0.77 \pm 0.21)
\end{equation}

where all the luminosities are in solar units and L$_{100\mu\textrm{m}}$ and L$_{160\mu\textrm{m}}$ are defined as $\nu L_{\nu}$. PACS-detected LAEs have measurements in 100$\mu$m and/or 160$\mu$m. For those detected in both bands, we obtain the $L_{IR}$ by using the calibration in the PACS-160$\mu$m band, since it is the closest one to the dust emission peak at our redshift ranges and, therefore, produces a more accurate result.

We can also obtain the total IR luminosities, i.e. dust content, for those PACS-undetected but MIPS-24$\mu$m detected SF GALEX-LAEs by using the \cite{Chary2001} calibrations between the total IR and 15$\mu$m and 12$\mu$m luminosities for local galaxies. These wavelengths are within the rest-frame ranges sampled through the MIPS-24$\mu$m band for LAEs at z$\sim$0.3 and $\sim$1.0, respectively. The applicability of these calibrations to high-redshift galaxies is discussed by \cite{Elbaz2010} who found that, below z$\sim$1.5, mid-IR extrapolations to the total IR luminosities are correct for SF galaxies below the (U)LIRG regime. The MIPS-24$\mu$m information was taken from the from S-COSMOS survey \citep{Sanders2007}. Only one object is not detected in that band. 

\begin{figure*}
   \centering
      \includegraphics[width=0.9\textwidth]{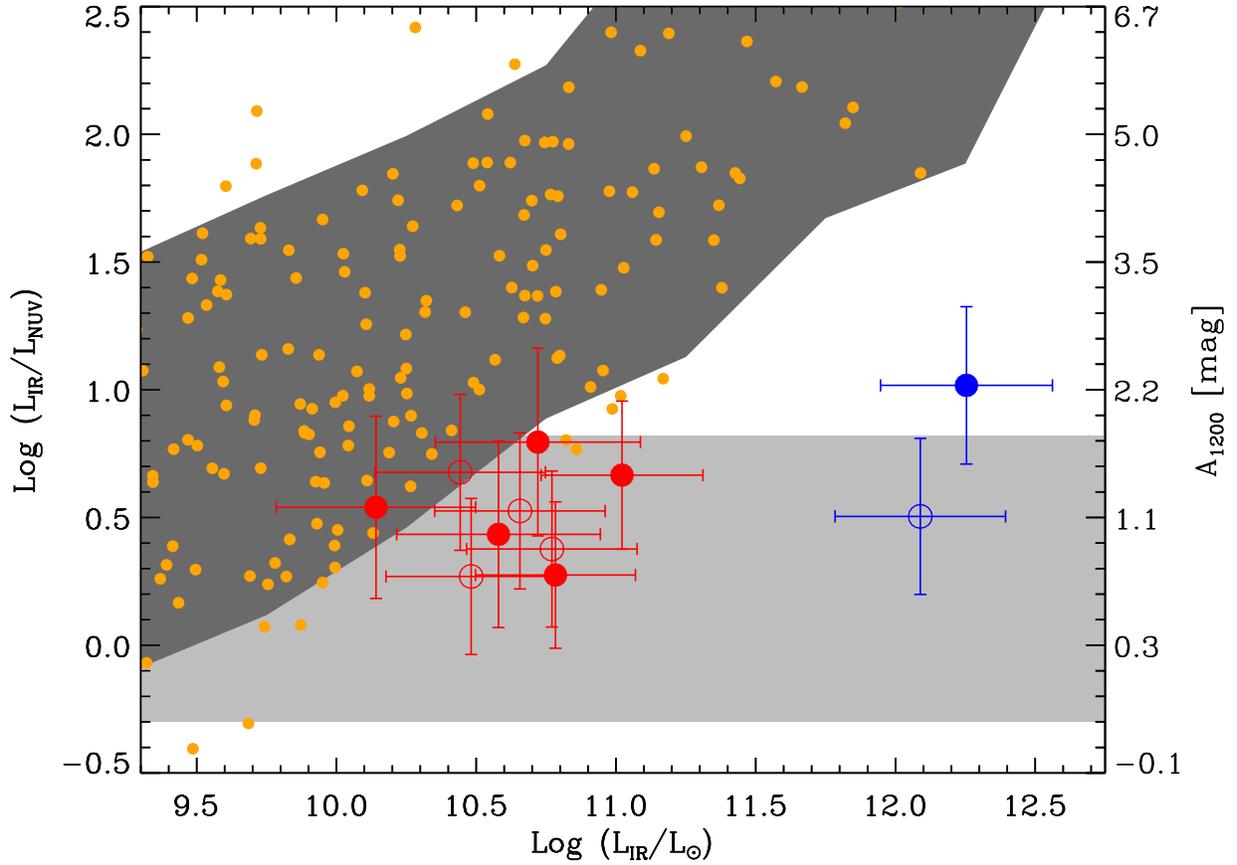}
   \caption{Ratio between total infrared and NUV luminosities against total infrared luminosity for star-forming GALEX-LAEs. Right vertical axis (dust attenuation) was built by using the calibration of \cite{Buat2005}. Yellow dots are data of nearby galaxies taken from \cite{Gil2007}. Dark shaded zone is the average region ($\pm$1.5$\sigma$) where these objects lay. Light shaded region corresponds to the dust content range of LAEs at z$\sim$0.3 reported by \cite{Finkelstein2009}. Red and blue dots are SF LAEs at z$\sim$0.3 and z$\sim$1.0, respectively. Filled symbols are PACS-dectected LAEs and open symbols are PACS-undetected LAEs. We do not plot the MIPS-24$\mu$m-undetected and PACS-undetected SF-LAE at z$\sim$0.3 (see text) given that there is no way to obtain its total IR luminosity with the calibrations used.
              }
           \label{buat}
   \end{figure*}

In Figure \ref{buat} we plot the ratio L$_{IR}$/L$_{UV}$ (or, equivalently, dust attenuation) versus L$_{IR}$ of our mid-IR/FIR detected GALEX-LAEs, besides the corresponding data of \cite[hereafter G07]{Gil2007} for a collection of nearby, SF galaxies. Our LAEs are located in the low-dust regime of the G07 data, indicating that objects exhibiting the Ly$\alpha$ line in emission at low and moderate redshift should not be as dusty as the whole population of galaxies at the same redshift. The dust attenuation in our mid-IR/FIR-detected LAEs at z$\sim$0.3 ranges within 0.75$\lesssim$$A_{1200\textrm{\AA}}$$\lesssim$1.8 mag, with a median value of 1.1 mag. This range broadens up to $A_{1200\textrm{\AA}}$ $\lesssim$2.5 when considering the z$\sim$1.0 LAEs as well. For the GALEX-LAE undetected both in MIPS-24$\mu$m and PACS we assume an upper limit in its dust attenuation equal to the minimum value obtained for the mid-IR/FIR detected GALEX-LAEs: $A_{1200\textrm{\AA}}\lesssim 0.75$. 

\cite{Finkelstein2011_espectros} studied dust content in LAEs at z$\sim$0.3 by using SED fitting and Balmer decrement. They derived an attenuation in the stellar continuum from Balmer decrement ranging from A$_{1200\textrm{\AA}}$$\sim$0 to $\sim$1.8, with a median value of 0.78 mag. These values are in agreement with those found via SED fitting, being most of them within the combined 1$\sigma$ uncertainties. \cite{Cowie2010} obtained dust attenuation from the NUV continuum slope and Balmer decrement of their LAEs at z$\sim$0.3, obtaining a median value of A$_{1200\textrm{\AA}}$$\sim$2.25 and $\sim$1.8, respectively, which are noticeably higher than those reported in \cite{Finkelstein2011_espectros}. \cite{Nilsson2009_letter} reported significantly higher values, most of them being A$_{1200\textrm{\AA}}$$\gtrsim$2.0, and as large as 7.5 mag. The results derived here are slightly higher than those derived in \cite{Finkelstein2011_espectros}, similar to \cite{Cowie2010} values, and lower than those in \cite{Nilsson2009_letter}. These discrepancies may be caused by the different methodologies employed, and even when they are same, the results do not agree quite well, likely due to the low number of objects in each sample.

The dust attenuation values reported up to z$\sim$2 are similar to those obtained in this work. However, at z$\gtrsim$3, while there are some measurements of high dust attenuation in high redshift LAEs, the vast majority of the reported extinction measurements are substantially lower than A$_{1200\textrm{\AA}}$$\lesssim$0.75 and are, in many cases, consistent with essentially no extinction. This could indicate that high redshift LAEs are less dusty than those at lower redshifts. However, note that, overall at the highest redshifts, the number of the studied LAEs at each redshift is low and, therefore, the results are not statistically significant. Furthermore, it is worth noting that SED fitting estimations (sometimes done with stacking analysis), spectroscopic results and IR measurements could produce systematic differences. This means that a possible evolution of dust attenuation with redshift should be taken with caution.



\section{Conclusions}\label{conclusion}

In this letter, we have reported the FIR detections of a sample of GALEX-selected LAEs at z$\sim$0.3 and z$\sim$1.0. Our main conclusions are as follows:

\begin{itemize}

 \item We find that a large fraction of LAEs at z$\sim$0.3 and z$\sim$1.0 are detected in the FIR. In the worst case, the fraction of FIR detected LAEs is about 15\%, although in most cases the fractions are higher than 30\%. Therefore, up to z$\sim$1.0, at least 30\% of the LAEs are expected to be detected in the FIR under the depth of the catalogs used. The percentages of MIPS-24$\mu$m detections are even higher, only one object being undetected in that band. 
 
 
 
 \item Based on direct measurements of the mid-IR/FIR SED, we conclude that the studied SF-LAEs at z$\sim$0.3 present a wide range of dust attenuation, 0.75$\lesssim$A$_{1200\textrm{\AA}}$$\lesssim$1.8, with  median value $A_{1200\textrm{\AA}}$ of 1.1. This range broadens up to 0.75$\lesssim$A$_{1200\textrm{\AA}}$$\lesssim$2.2 when considering LAEs at z$\sim$1.0. We only find one object, undetected both in MIPS-24$\mu$m and PACS, whose dust content has an upper limit of $A_{1200\textrm{\AA}}\lesssim$ 0.75. 
 
 
\item The dust attenuation values derived here are comparable to those reported in previous works up to z$\sim$2.0, pointing out that there is no significant evolution in dust content in that epoch. At z$\gtrsim$3, although some works showed the existence of dusty LAEs, most of the extinction measurements are lower  than those at z$\lesssim$2.0, which could indicate an evolution with redshift. However, the different methods used to derive the results and the low number of LAEs at certain redshifts make the results statistically unsignificant.

\end{itemize}

\acknowledgements

We thank the anonymous referee for providing helpful comments which have improve the quality of this paper. This work was supported by the Spanish Plan Nacional de Astrononom\'ia y Astrof\'isica under grant AYA2008-06311-C02-01. Some/all of the data presented in this Letter were obtained from the Multimission Archive at the Space Telescope Science Institute (MAST). STScI is operated by the Association of Universities for Research in Astronomy, Inc., under NASA contract NAS5-26555. Support for MAST for non-HST data is provided by the NASA Office of Space Science via grant NNX09AF08G and by other grants and contracts {\it Herschel} is an ESA space observatory with science instruments provided by European-led Principal Investigator consortia and with important participation from NASA. The \emph{Herschel} spacecraft was designed, built, tested, and launched under a contract to ESA managed by the Herschel/Planck Project team by an industrial consortium under the overall responsibility of the prime contractor Thales Alenia Space (Cannes), and including Astrium (Friedrichshafen) responsible for the payload module and for system testing at spacecraft level, Thales Alenia Space (Turin) responsible for the service module, and Astrium (Toulouse) responsible for the telescope, with in excess of a hundred subcontractors. PACS has been developed by a consortium of institutes led by MPE (Germany) and including
UVIE (Austria); KUL, CSL, IMEC (Belgium); CEA, OAMP (France); MPIA (Germany); IFSI, OAP/AOT, OAA/CAISMI,
LENS, SISSA (Italy); IAC (Spain). This development has been supported by the funding agencies BMVIT (Austria), ESA-
PRODEX (Belgium), CEA/CNES (France), DLR (Germany), ASI (Italy) and CICYT/MICINN (Spain).

\clearpage

\clearpage

\begin{table*}
\begin{center}
\caption{Summary of some previous results in dust attenuation in LAEs. Reference, redshift, dust attenuation range and the method followed to obtain it are shown: SED fittings with BC03 templates to the individual or stacked observed photometry (TF-BC03 and TF-BC03 \& Stacked, respectively), optical spectroscopic analysis (Spec.), UV continuum slope and mid-IR measurements.\label{dust_sum}}
\begin{tabular}{c c c c}
\tableline\tableline
Reference & redshift & Dust attenuation [mag] & Method\\ \hline
\tableline
\cite{Finkelstein2009} & $\sim$0.3 & 0.0 $\lesssim$ $A_{1200\AA}$ $\lesssim$ 2.0 & TF-BC03 \\
\cite{Finkelstein2011_espectros} & $\sim$0.3 & 0.4 $\lesssim$ $A_{1200\AA}$ $\lesssim$ 1.6 & Spec \\
\cite{Nilsson2009_letter} & $\sim$0.3 & 2.0$\lesssim$ $A_{1200\AA}$ $\lesssim$ 7.5 & IRAC-8$\mu$m measurements \\
\cite{Cowie2010} & $\sim$0.3 & A$_{1200\AA}$ $\sim$ 2.25 & UV continuum slope \\
\cite{Cowie2010} & $\sim$0.3 & A$_{1200\AA}$ $\sim$ 1.8 & Spec \\
\cite{Guaita2011} & $\sim$2.1 & 0.0$\lesssim$ $A_{1200\AA}$ $\lesssim$ 2.41 & TF-BC03 \\
\cite{Nilsson2011} & $\sim$2.3 & 0.0$\lesssim$ $A_{1200\AA}$ $\lesssim$ 10.0 & TF-BC03 \\
\cite{Nilsson2009_letter} & $\sim$2.3 & 4.0$\lesssim$ $A_{1200\AA}$ $\lesssim$ 8.0 & MIPS-24$\mu$m measurements \\
\cite{Gawiser2007} & $\sim$3.1 &  A$_{1200\AA}$ $\lesssim0.5$ & TF-BC03 \& Stacked \\
\cite{Lai2008} & $\sim$3.1 &  A$_{1200\AA}$ $\sim0.0$ & TF-BC03 \& Stacked \\
\cite{Ono2010} & $\sim$3.1 &  A$_{1200\AA}$ $\sim0.25$ & TF-BC03 \& Stacked \\
\cite{Ono2010} & $\sim$3.1 & 0.0$\lesssim$ $A_{1200\AA}$ $\lesssim$ 5.6 & TF-BC03 \\
\cite{Nilsson2007} & $\sim$3.15 &  A$_{1200\AA}$ $\sim 0.7$ & TF-BC03 \& Stacked \\
\cite{Ono2010} & $\sim$3.7 &  A$_{1200\AA}$ $\sim1.5$ & TF-BC03 \& Stacked \\
\cite{Ono2010} & $\sim$3.7 & 0.3$\lesssim$ $A_{1200\AA}$ $\lesssim$ 5.4 & TF-BC03 \\
\cite{Finkelstein2008} & $\sim$4.4 & 0.4$\lesssim$ $A_{1200\AA}$ $\lesssim$ 1.8 & TF-BC03 \\
\cite{Finkelstein2009d} & $\sim$4.5 & 0.3$\lesssim$ $A_{1200\AA}$ $\lesssim$ 4.5 & TF-BC03 \\
\cite{Pirzkal2007} & $\sim$5.0 & 0.0$\lesssim$ $A_{1200\AA}$ $\lesssim$ 1.8 & TF-BC03\\
\cite{Lai2007} & $\sim$5.7 & 1.20$\lesssim$ $A_{1200\AA}$ $\lesssim$ 1.81 & TF-BC03\\

\tableline
\end{tabular}
\end{center}
\end{table*}

\begin{table}
\begin{center}
\caption{Number of GALEX-selected LAEs within the surveyed area of PACS in COSMOS and GOODS-South fields. We separate the objects in redshift, nature (according to the UV spectra, see text) and the field where they are located. Numbers between parentheses are those detected in the FIR with PACS.\label{table_number}}
\begin{tabular}{l c c c}
\\
\tableline\tableline
Field & Type & z$\sim$0.3 & z$\sim$1.0 \\ \hline
COSMOS & AGN & 3 (1) & 14 (2) \\
COSMOS & SF & 9 (4) & 2 (1) \\
\hline
GOODS-South & AGN & 0 (0) & 3 (2) \\
GOODS-South & SF & 1 (1) & 0 (0) \\
\tableline
\end{tabular}
\end{center}
\end{table}

\end{document}